# Operation of MHSP multipliers in high pressure pure noble-gas

F. D. Amaro[a], J. F. C. A. Veloso[a,b], A. Breskin[c], R. Chechik[c], J. M. F. dos Santos[a,*]

[a]Physics Dept., University of Coimbra, 3004-516 Coimbra, Portugal
[b]Physics Dept., University of Aveiro, 3810-193 Aveiro, Portugal
[c]Dept. of Particle Physics, The Weizmann Institute of Science, 76100 Rehovot, Israel

## Abstract

We report on the performance of a Micro-Hole & Strip Plate (MHSP) electron multiplier operating in pure Xe, Kr, Ar and Ne at the pressure range of 1 to 6 bar. The maximal gains at 1 bar Xe and Kr are $5 \times 10^4$ and $10^5$, respectively; they drop by about one order of magnitude at 2 bar and by almost another order of magnitude at 5-6 bar; they reach gains of 500 and 4000 at 5 bar in Xe and Kr, respectively. In Ar, the gain varies very little with pressure, being $3-9 \times 10^3$; in Ne the maximum attainable gain, about $10^5$, is pressure independent above 2 bar. The results are compared with that of single- and triple-GEM multipliers operated in similar conditions. Potential applications are in hard X-ray imaging and in cryogenic radiation detectors.

**Submitted to *JINST*, Jan 9, 2006**

# 1. Introduction

Particular interest has been given lately to gas avalanche electron multipliers operating in noble gases and noble-gas mixtures. Relevant applications of such devices are in the fields of cryogenic detectors for neutrino physics, dark matter search and PET [1-4], gas photomultipliers [5], X-ray and neutron imaging [6,7] etc. The very simple high-purity noble-gas handling and their low aging under gas avalanche permit the design of sealed detectors with stable, long-term operation. However, charge multiplication in these media has been usually strongly limited by photon- and ion-mediated secondary processes, which motivated recent studies to overcome this drawback.

Recently, progress has been made in high-gain operation of multi-GEMs (Gas Electron Multipliers) in noble gases and their mixtures, over a broad pressure range [7-10]. The avalanche confinement within the GEM holes effectively hinders photon-mediated secondary processes, allowing for reaching high gains even in highly UV-emissive gases [11,12]. In particular, detailed studies have been carried out to investigate the performance of single- and triple-GEM multipliers operated in high-pressure noble gases [8,13].

A noticeable drawback of micropattern detectors such as Micro Strip Gas Chambers (MSGCs) and GEMs is the gain drop at elevated gas pressure, which is mainly due to the limited voltage that can be applied to the microstructure device before the discharge limit, and partly due to secondary photon- and ion-feedback effects. The only exception occurs with the light noble gases, helium and neon, where high gains were achieved at high pressures [8]. With a triple-GEM detector operating in heavy noble gases, Kr and Xe, a gain drop of four orders of magnitude was observed when increasing the pressure from 1 to 5 bar [13], and gains below 10 were recorded at 5 bar. Gains of 500 to 80 were measured in pure Kr, using a single-GEM multiplier, for respective pressures of 1 to 10 bar [8].

The recently proposed Micro-Hole & Strip Plate electron multiplier (MHSP) [14-16] seems to be very adequate for the operation in high-pressure noble gases. It combines GEM-like and MSGC-like multiplying elements in a single structure, resulting in two successive multiplication stages: hole multiplication followed by anode-strip multiplication. The MHSP has good optical screening of avalanche photons and improved ion-blocking capability as compared to GEM [16]. It has fast signals [17] and good localization properties [15]; it can operate as a single-element detector or as a final amplifying element in a cascaded multiplier [15,16].

The first attempts to operate a MHSP in Ar/50 mbar Xe mixtures at pressures between 1 to 7 bar, yielded maximal gains of about $10^3$ at 7 bar [18]. The maximum achievable gain did not show significant pressure dependence up to 6 bar, above which there was a sharp drop, probably due to some discharge limits. Energy resolutions between 14% and 16% FWHM were recorded with 6 keV X-rays, up to 6 bar.

In this work we report on the performance of a single-MHSP multiplier, operating in pure noble gases, Ne, Ar, Kr and Xe, in a pressure range of 1-6 bar. The results are compared with those obtained with single- and triple-GEM multipliers. Similarly to the results reported with triple-GEMs, the gain dependence on pressure is affected by the gas composition; higher gains were attained at high pressures in the lighter noble gases [13].

## 2. Detector description

The detector is located within a stainless steel vessel, 10 cm in diameter, having a 25-$\mu$m thick aluminized Mylar window, 2-mm in diameter, glued to the stainless steel with a low vapour-pressure epoxy (Trac-Con 2116). A Macor structure supports the MHSP foil and carries the electrical contacts to the MHSP electrodes. The various voltages were supplied through feedthroughs, glued to the stainless steel vessel with the same epoxy resin. The absorption/drift region and the induction region gaps, above and below the MHSP multiplier, respectively, are 5 and 3 mm wide. The detector was vacuum pumped ($10^{-5}$ mbar) and, then, filled with noble gases at different pressures without baking; it was sealed off from the vacuum/gas-filling system during the measurements. The gas purity was maintained using non-evaporable getters (SAES St707), heated at about 150ºC and placed in a small annex volume connected to the main detector volume. No gas aging was noticed trough all the experiments.

The MHSP electrode, with an active area of 28 x 28 mm$^2$, is made of a 50 $\mu$m thick Kapton with a 5 $\mu$m copper clad coating on both sides. The top surface has a GEM like pattern with bi-conical holes of about 40/70 $\mu$m in diameter, arranged in an asymmetric hexagonal lattice of 140- and 200-$\mu$m pitch in the direction parallel and perpendicular to the strips pattern in the bottom side, with the holes centred within the cathode strips, ~100 $\mu$m wide, while the anodes, ~35 $\mu$m wide, run between them, in a 200 $\mu$m pitch (see photograph in ref [17]).

Within the present studies, the detector was irradiated with 5.9 keV X-rays from a [55]Fe X-ray source with the 6.4 keV X-rays filtered by a chromium film. The primary electron cloud resulting from the 5.9 keV X-ray interactions in the drift region are focused into the holes where they undergo avalanche multiplication (Fig.1). The avalanche electrons are extracted out of the holes towards the anode strips where they are further multiplied and collected. The signals from the MHSP anode strips were fed through a Canberra 2006 preamplifier of 1.5 V/pC; for the Ne measurements the sensitivity was reduced to 0.3 V/pC. The signals were further processed by a Tennelec TC243 amplifier (4 μs shaping time) and a Nucleus PCA2 1024 multichannel analyser. The electronic chain sensitivity was calibrated, for absolute gain determination, using a calibrated capacitor directly connected to the preamplifier input and to a precision pulse generator.

All the electrodes were independently polarized. The detector vessel, the radiation window and the induction backplane were grounded. The voltage of the MHSP top electrode, $V_{TOP}$, determines the drift field; the voltage difference between this top electrode and the cathode strips, $V_{C-T}$, determines the avalanche gain in the holes; the voltage difference between the anode and cathode strips, $V_{A-C}$, determines the avalanche gain around the anode strips; and the voltages of the strips determine the induction field.

## 3. Experimental results and discussion

In each set of measurements at a given pressure, a maximum voltage across the holes, $V_{C-T}$, and across the strips, $V_{A-C}$, was established. Different combinations of these two values were experimented in order to maximize the gain, without reaching the onset of discharge. $V_{A-C}$ was, then, gradually decreased, while keeping $V_{C-T}$ constant. The 5.9 keV X-rays pulse-height distributions were fitted to a Gaussian superimposed on a linear background and the peak centroid was monitored as function of $V_{A-C}$. For each gas, $V_{TOP}$ was increased with pressure so that the reduced electric field in the drift region was only mildly decreased. Values in the range of 100-75 $Vcm^{-1}bar^{-1}$ were used for Xe, Kr and Ar fillings, and of 60-50 $Vcm^{-1}bar^{-1}$ for Ne fillings, respectively.

In Fig.2 *a-d*, we present the detector's total gain as function of the total voltage difference applied to the MHSP, $V_{Total}=V_{C-T} + V_{A-C}$, for pure Xe, Kr, Ar and Ne, respectively, and for the different gas pressures. The gain-curves exhibit the characteristic exponential avalanche growth, but at low $V_{Total}$ values the pulse amplitudes drop faster than exponential, due to inefficient electron transport to the anode strips. Fixing $V_{A-C}$ and varying $V_{C-T}$, instead, the slope of the exponential variation of the gain is different but the maximal total gain achieved is the same. The $V_{C-T}$ values used for each pressure, were increased with increasing pressure from 460 to 820V for Xe, from 430 to 740V for Kr and from 320 to 660V for Ar; for Ne we set values around 320V for all the pressures, except 260V set at 1 bar.

Fig. 2*a,b*, show an identical trend of the maximum gain dependence on pressure for pure Xe and Kr: a fast decrease of the maximum achievable gain from 1 to 2 bar and a slower decrease of this maximum for pressures above 2 bar. Nevertheless, the amplitude reduction with increasing pressure is much slower for Kr than for Xe. Gains about $5x10^4$ and $10^5$ were obtained at 1 bar for Xe and Kr, respectively, being reduced to $5x10^3$ and $2x10^4$ at 2 bar, and to 500 and $4x10^3$ at 5 bar. On the other hand, the maximum gain obtained for Ar (Fig.2*c*) presents only a small dependence on the pressure, increasing from $\sim5x10^3$ at 1 bar to a maximum of $10^4$ at 4 bar and decreasing by a factor 3 at 6 bar. The maximum gain achieved in Ne is fairly constant for pressures above 1 bar, being $\sim2x10^4$ at 1 bar and $\sim10^5$ for all the other gas pressures.

Fig.3 summarizes the maximum gain achieved in the MHSP detector as a function of gas pressure for the different noble gases. For comparison, we include the maximum gains achieved with a triple-GEM [13] and a single-GEM in Kr [8].

Compared to the triple-GEM multiplier, the MHSP presents a much slower decrease of the maximum gain with increasing pressure, in Xe, Kr and Ar. The main reason for this difference is related to the total voltage that can be applied to the multipliers as the pressure increases.

The maximum operation voltage that can be applied to the MHSP is presented in Fig.4 as a function of the pressure for the different gas fillings, together with the maximum voltage that can be applied across each GEM, in a triple-GEM mode [13] and in a single GEM, in Kr [8]. While in the MHSP the maximum applicable total voltage steadily increases with pressure, from values around 600 to around 1100V for Xe, Kr and Ar, the maximum voltage difference applicable across each GEM, in a triple-GEM cascade, saturates when the pressure increases above 2, 3 and 4 bar for Xe, Kr and Ar, respectively [13]. This last effect is attributed to ion-induced electron emission, occurring in noble gases, due to ion feedback from the last to the preceding GEMs [7], which limits the maximum applicable voltage [13]. This effect is considerably reduced in single-element multipliers such as single-GEM [8] and MHSP. On the other hand, for high pressure Ne, the maximum applicable voltage is fairly pressure-independent, as in triple-GEM operating in Ne and He [13].

Studies on the electron avalanche mechanisms have been recently performed [13,19,20]. For Xe, Kr and Ar the electron avalanche ionisation is determined by the electron-impact mechanism, which explains the maximum gain drop for high pressures as the maximum applied voltage does not increase as fast as pressure [13]. For dense light noble gases other mechanisms, such as associative ionisation and/or penning ionisation with impurities, predominate over electron-impact ionisation [13,19,20].

## 4. Conclusions

In this work we presented the characteristics of a MHSP electron multiplier operated in Xe, Kr, Ar and Ne at pressures ranging from 1 to 6 bar. It was shown that in most cases, this single-element multiplier yielded higher gains than those reached with single- and triple-GEM elements. This could originate from the particular MHSP's geometry; the two amplification stages separated by only a few tens of microns, resulting in a more efficient electron transfer from stage to stage compared to that occurring in multi-GEM cascades [21].

The gas gain at 1 bar is about $5 \times 10^4$ for Xe, and higher than $10^5$ for Kr. Xe and Kr show fast gain decay with pressure and the gains are reduced to about 500 and 4000, respectively, at 5 bar. For Ar, the gain variation with pressure is less marked; it varies between 3 and $9 \times 10^3$, with a maximum at 4 bar. In Ne, the maximum achievable gain increases from $2 \times 10^4$ at 1 bar to around $10^5$ for pressures above 2 bar.

At atmospheric pressure, the values reached with a MHSP are somewhat higher compared to a triple-GEM, except for Ar where the gains are lower by almost one order of magnitude. For pressures above 4 bar the MHSP reaches about two orders of magnitude higher gains than the triple-GEM. In Ne, however, the gain difference between the two multipliers is reduced with increasing pressure; it becomes similar above 5 bar.

Compared to a single-GEM operated in pure Kr, the MHSP yields gains that are more than two orders of magnitude higher, at 1 bar; the MHSP's maximum gain decreases faster with increasing pressure, resulting in only one order of magnitude gain difference between the two at 6 bar.

**Acknowledgements**

This work was supported in part by Project POCTI/FNU/50360/02 through FEDER and FCT (Lisbon) programs and by the Israel Science Foundation project 151/01. A. Breskin is the W.P. Reuther Professor of Research in peaceful use of atomic energy. We acknowledge the efforts of Rui de Oliveira, EST/DEM/PMT-CERN in improving the MHSP quality.

**Figure Captions**

Fig.1-Shematic diagram of the MHSP detector and its operating principle.

Fig.2 – MHSP gain as a function of the total voltage difference applied to the MHSP, $V_{Total} = V_{C-T} + V_{A-C}$, for xenon (*a*), krypton (*b*), argon (*c*) and neon (*d*) and for different filling pressures.

Fig.3 – Maximum gain of different detectors as a function of gas pressure for the different noble gases: solid lines - MHSP detector [this work]; broken lines - triple-GEM detector [13] and single-GEM detector in Kr [8].

Fig.4 – Maximum operation voltage as a function of the pressure for the different gas fillings: solid lines – total voltage, $V_{Total}$ applied across the MHSP multiplier [this work]; broken lines – total voltage applied across each GEM, in a triple-GEM [13] and in a single-GEM in Kr [8].

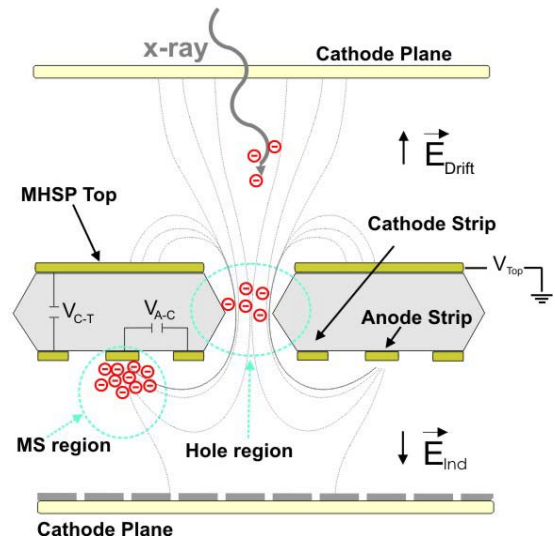

Fig.1-Shematic diagram of the MHSP detector and its operating principle.

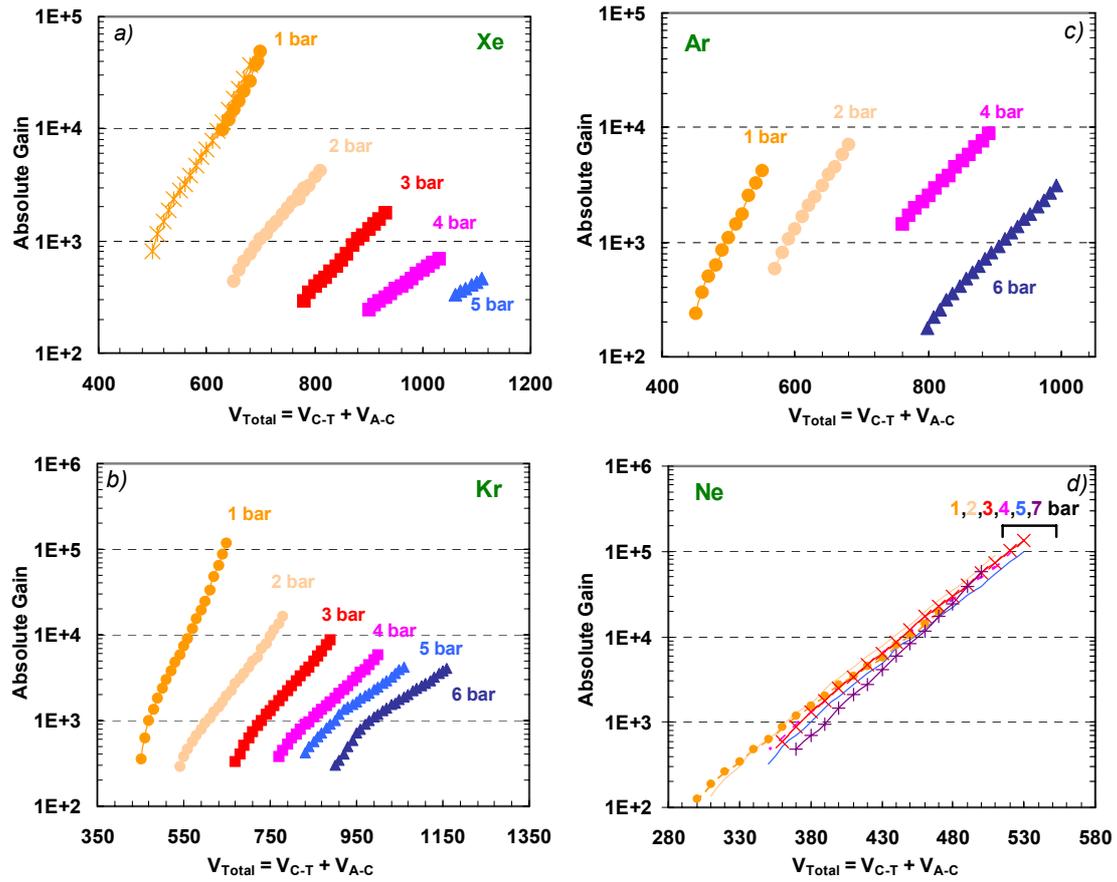

Fig.2 – MHSP gain as a function of the total voltage difference applied to the MHSP, $V_{Total} = V_{C-T} + V_{A-C}$, for xenon (*a*), krypton (*b*), argon (*c*) and neon (*d*) and for different filling pressures.

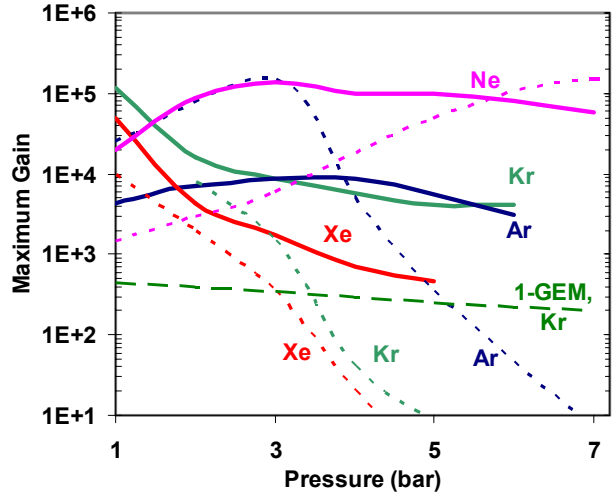

Fig.3 – Maximum gain of different detectors as a function of gas pressure for the different noble gases: solid lines - MHSP detector [this work]; broken lines - triple-GEM detector [13] and single-GEM detector in Kr [8].

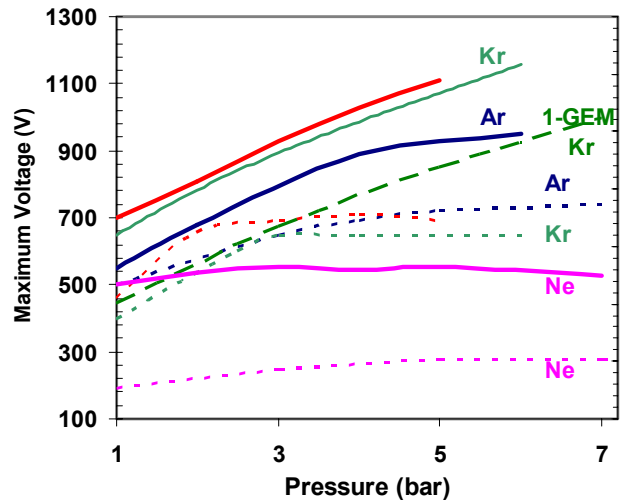

Fig.4 – Maximum operation voltage as a function of the pressure for the different gas fillings: solid lines – total voltage, $V_{Total}$ applied across the MHSP multiplier [this work]; broken lines – total voltage applied across each GEM, in a triple-GEM [13] and in a single-GEM in Kr [8].